\documentclass[aps,prb,reprint,longbibliography,superscriptaddress]{revtex4-2}

\usepackage{graphicx}
\usepackage{amsmath}

\DeclareMathOperator{\real}{Re}
\DeclareMathOperator{\imag}{Im}

\begin{document}

\title{Squeezing anyons for braiding on small lattices}

\author{N. S. Srivatsa}
\thanks{These authors contributed equally to this work.}
\affiliation{Max-Planck-Institut f\"{u}r Physik komplexer Systeme, D-01187 Dresden, Germany}
\author{Xikun Li}
\thanks{These authors contributed equally to this work.}
\affiliation{Max-Planck-Institut f\"{u}r Physik komplexer Systeme, D-01187 Dresden, Germany}
\affiliation{School of Physics and Materials Science, Anhui University, Hefei, Anhui 230601, China}
\author{Anne E. B. Nielsen}
\affiliation{Max-Planck-Institut f\"{u}r Physik komplexer Systeme, D-01187 Dresden, Germany}
\affiliation{Department of Physics and Astronomy, Aarhus University, 8000 Aarhus C, Denmark}

\begin{abstract}
Adiabatically exchanging anyons gives rise to topologically protected operations on the quantum state of the system, but the desired result is only achieved if the anyons are well separated, which requires a sufficiently large area. Being able to reduce the area needed for the exchange, however, would have several advantages, such as enabling a larger number of operations per area and allowing anyon exchange to be studied in smaller systems that are easier to handle. Here, we use optimization techniques to squeeze the charge distribution of Abelian anyons in lattice fractional quantum Hall models, and we show that the squeezed anyons can be exchanged within a smaller area with a close to ideal outcome. We first use a toy model consisting of a modified Laughlin trial state to show that one can shape the anyons without altering the exchange statistics under certain conditions. We then squeeze and braid anyons in the Kapit-Mueller model and an interacting Hofstadter model by adding suitable potentials. We consider a fixed system size, for which the charge distributions of the normal anyons overlap, and we find that the outcome of the exchange process is closer to the ideal value for the squeezed anyons. The time needed for the exchange is also important, and for a particular example we find that the duration needed for the process to be close to the adiabatic limit is about five times longer for the squeezed anyons when the path length is the same. Finally we show that the exchange outcome is robust with respect to small modifications of the potential away from the optimized value.
\end{abstract}

\maketitle

\section{Introduction}

Topologically ordered phases of matter are breeding grounds for anyonic quasiparticles with exchange statistics that are neither fermionic nor bosonic \cite{Leinaas,wilczek}. The exchange statistics are robust against local noise, which has motivated much work towards utilizing anyons for quantum computing \cite{TopQC}. Anyons appear, e.g., in fractional quantum Hall systems which are two-dimensional electronic systems subject to strong magnetic fields \cite{Laughlin,halperin,arovas}. Several lattice models hosting fractional quantum Hall physics and anyons have also been found \cite{Sorensen,Hafezi,He,Neupert,Hormozi,Ra,kapit2010}. The search for lattice models is, in part, motivated by the interest in realizing fractional quantum Hall physics in ultracold atoms in optical lattices, which would allow for detailed investigations of the effect, even at the level of single particles.

Anyons can be trapped at specific positions with pinning potentials \cite{Storni,johri}. By adiabatically moving these potentials one can braid the anyons and in this way get access to the anyonic braiding transformations \cite{kapit2012,yangle,Liu,Jaworowski}. To get the topological braiding properties, however, the anyons need to be well separated, and since the anyons are quasiparticles that are spread out over a region, this means that the system needs to have a certain size.

The requirement to have sufficiently large system sizes is, however, challenging. From a theoretical point of view, the computational resources needed to study a quantum many-body system in general grows exponentially with system size, and this is often a limitation for studying braiding statistics. From an experimental point of view, it is generally challenging to keep coherences in large quantum systems, and dealing with small system sizes is also an advantage for ultracold atoms in optical lattices \cite{Ra}. From a computational point of view, the size requirements put a limit on the number of qubits per area. It would hence be helpful, if one could reduce the area needed for braiding.

Here, we show that one can reduce the area needed for braiding by using optimal control techniques to shape the anyons in such a way that overlap between the charge distributions of the anyons is avoided. We first consider a toy model, which consists of a modified Laughlin type trial state on a lattice. Since the braiding statistics is a topological quantity, it should be possible to make local deformations to shape the anyons without altering the braiding statistics, and we show how this comes about in the toy model under certain conditions. We then consider the Kapit-Mueller model and an interacting Hofstadter model as examples of lattice fractional quantum Hall models. We choose open boundary conditions, since this is the most relevant case for experiments. The shaping of the anyons is done with a position dependent potential. For the system size considered, the anyons overlap significantly when they are created from a local potential, and the phase acquired when two anyons are exchanged differs from the ideal value predicted for well separated anyons. Using the optimized potential removes the overlap of the charge distributions of the anyons, and the phase acquired when exchanging two anyons is significantly closer to the ideal value.

The modifications done to shape the anyons could affect the size of the gap and the ability to couple to the excited states, and we therefore also compute the time needed for doing the exchange process in order to be close to the adiabatic limit. For the example considered, we find that this time is about a factor of five bigger for the case with the optimized potential compared to the case with the local potentials. With the optimized potential, we hence need to do the operation more slowly to get the improvement in the exchange phase. If we instead use the local potentials, we would need to increase the system size to improve the results. This is expected to also lead to an increased duration, due to the increase in the length of the path.

It is important to note that the quantity we optimize is the charge distribution of the anyons and not the statistical phase itself. With the method used, the topological robustness remains in the model. We show numerically that if the potential is modified slightly away from the optimal choice, the phase acquired when exchanging two anyons in the considered finite size system remains practically the same.

The paper is structured as follows. In Sec.~\ref{sec:wf}, we introduce the toy model for squeezed anyons and compute the braiding statistics. In Sec.~\ref{sec:opt}, we consider the Kapit-Mueller model and an interacting Hofstadter model. We explain how we squeeze and exchange the anyons, and we compute the improvement in the exchange phase. We also study the robustness of the exchange phase with respect to slight modifications of the potential away from the optimal choice, and we estimate how slowly the anyons need to move to be close to the adiabatic limit. Section \ref{sec:conclusion} concludes the paper.

\section{A toy model}\label{sec:wf}

Topological properties are robust against local deformations, as long as the deformations are not so large that they bring the system out of the topological phase or allow the topological quantity in question to switch to one of the other allowed values. We therefore expect that it should be possible to shape the anyons to some extent, while keeping the braiding properties unaltered. We start out by showing this explicitly for a simple model, namely a modified Laughlin trial state on a lattice. This system can be analyzed using a combination of analytical observations and Monte Carlo simulations, and this allows us to study large systems with well-separated anyons.

\subsection{Lattice Laughlin states}

We first consider the case without shaping. It is well-known how one can modify the Laughlin state \cite{Laughlin} with quasiholes defined on a disk-shaped region in the two-dimensional plane into a Laughlin state with quasiholes defined on a square lattice with open boundary conditions \cite{anne2015}. This is done by restricting the allowed particle positions to the lattice sites and by also restricting the magnetic field to only go through the lattice sites. Here, we consider a system with one magnetic flux unit through each lattice site, and we take the charge of a particle to be $-1$. The resulting wavefunction with $S$ quasiholes at the positions $w_i$, with $i=1,2,\ldots,S$, is given by \cite{anne2015}
\begin{equation}
|\psi_{q,S} \rangle = \sum_{n_1,n_2,\ldots,n_N}
\psi_{q,S}(n_1,n_2,\ldots,n_N)
|n_{1},n_{2},\ldots,n_{N}\rangle,
\end{equation}
where
\begin{multline}\label{wf1}
\psi_{q,S}(n_1,n_2,\ldots,n_N)=
C^{-1}\,\delta_{n}\,
\prod_{j,k}(w_j-z_k)^{p_jn_k}\\
\times\prod_{i<j}(z_{i}-z_{j})^{q n_{i} n_{j}}
\prod_{i\neq j}(z_{i}-z_{j})^{-n_{i}}.
\end{multline}
In this expression, the $z_i$, with $i=1,2,\ldots,N$, are the coordinates of the $N$ lattice sites written as complex numbers, $n_j\in\{0,1\}$ is the number of particles on the $j$th site, $q$ and $p_j$ are not too large integers, and $q$ must be at least $2$. $C$ is a real normalization constant that depends on all the $w_i$, and
\begin{equation}\label{delta}
\delta_{n} =
\begin{cases}
  1 &   \textrm{for }\sum^{N}_{j=1}n_j=\left(N-\sum^{S}_{j=1}p_j\right)/q\\
  0 &   \text{otherwise}
\end{cases}
\end{equation}
fixes the number of particles $\sum_j n_j\gg 1$ in the system in such a way that there are $q$ flux units per particle and $p_k$ flux units per quasihole. The particles are fermions for $q$ odd and hardcore bosons for $q$ even. We restrict $q$ and $p_j$ to small integers and require the number of particles to be large compared to one, since this is the regime for which it has already been confirmed numerically \cite{anne2015,ivan2016} that the lattice Laughlin state has the desired topological properties.

When the state is topological, one observes the following results numerically \cite{anne2015,ivan2016}. The state without anyons has a uniform density of $1/q$ particles per site in the bulk of the system, which means that $\langle \psi_{q,0}|n_j|\psi_{q,0}\rangle$ is constant and equal to $1/q$ in the bulk. The $k$th quasihole creates a local region around $w_k$ with a lower particle density. This can be quantified by considering the density difference
\begin{equation}\label{densdiff}
\rho(z_j)=\langle\psi_{q,S}| n_j|\psi_{q,S}\rangle
-\langle \psi_{q,0}| n_j|\psi_{q,0}\rangle,
\end{equation}
which is defined as the expectation value of $n_j$ in the state with anyons minus the expectation value of $n_j$ in the state without anyons. When all other anyons are far away, the total number of particles missing in the local region is $p_k/q$. This must be so when the quasihole is entirely inside the local region, since it follows from Eq.\ \eqref{delta} that the presence of the $k$th quasihole reduces the number of particles in the system by $p_k/q$. Since the particles have charge $-1$, the quantity $-\rho(z_j)$ is the charge distribution of the anyons, and summing $-\rho(z_j)$ over the local region at $w_k$ gives the charge $p_k/q$ of the $k$th anyon.

As an example, we plot $\rho(z_j)$ for the state $|\psi_{2,2}\rangle$ with two quasiholes of charge $1/2$ in Fig.\ \ref{fig:shape}(a). Each of the quasiholes occupies a roughly circular region and spreads over about 16 lattice sites. We obtain $-\sum_{j\in B}\rho(z_j)\approx 0.5$ as expected, where $B$ is the set of the labels of the $16$ sites inside one of the dashed circles. Figure \ref{fig:shape}(b) shows $\rho(z_j)$ along a line parallel to the real axis that goes through one of the quasiholes.

\subsection{Shaping}

The idea is to shape the quasiholes by splitting the factors $(w_j-z_k)$ appearing in the wavefunction into several pieces. The splitting gives us additional parameters that we can tune to obtain the desired shape. We first study how the splitting affects the density difference, and in the next section we will show under which conditions the shaped objects have the same braiding statistics as the original anyons. Specifically, we define the $L$ weights $g_j^{h}$ that are real numbers and the $L$ coordinates $w_j^{h}$ that are complex numbers. The weights fulfil the constraint $\sum_{h=1}^{L}g_k^{h}=p_k$. The resulting state takes the form
\begin{multline}\label{wf2}
\psi_{q,S}(n_1,n_2,\ldots,n_N)=
C^{-1}\,\delta_{n}\,
\prod_{h,j,k}(w_j^{h}-z_k)^{g_j^{h}n_k}\\
\times\prod_{i<j}(z_{i}-z_{j})^{q n_{i} n_{j}}
\prod_{i\neq j}(z_{i}-z_{j})^{-n_{i}}.
\end{multline}
The state \eqref{wf1} is the special case with $L=1$.

\begin{figure}
\includegraphics[width=\columnwidth]{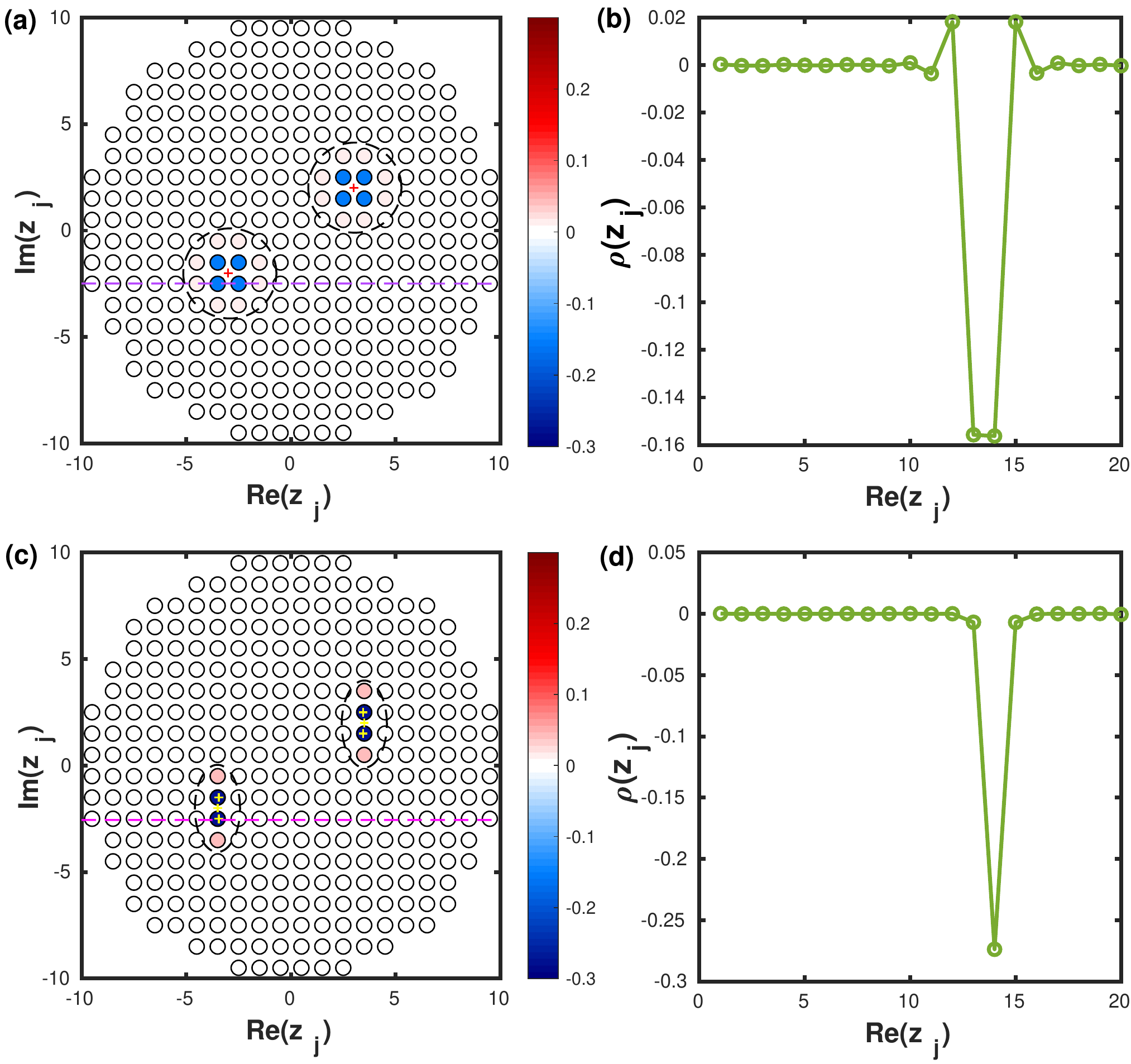}
\caption{(a) Density difference $\rho(z_j)$ [see Eq.\ \eqref{densdiff}] of two quasiholes with charge 1/2 in the state \eqref{wf1}. The quasihole positions $w_k$ are depicted by red plus symbols. The charge has a spread of roughly $16$ lattice sites (the sites that fall within the dashed circles). (b) Density difference along the row marked by the purple line in (a). (c) Density difference of squeezed quasiholes of charge 1/2 constructed by using three weights with $g_1^h=g_2^h=1/3$ (at the positions indicated by the green and yellow plus symbols) using Eq.~\eqref{wf2}. The charge has a spread of roughly 8 lattice sites (the sites that fall within the dashed ellipses). (d) The density difference on the lattice sites along the row marked by a purple line in (c) also shows the squeezing. In both cases, $q=2$, the number of lattice sites is $N=316$, and we choose a square lattice with a roughly circular edge to mimic a quantum Hall droplet.}\label{fig:shape}
\end{figure}

We show one example in Fig.\ \ref{fig:shape}(c), where we plot the density profile for the state \eqref{wf2} with $q=2$, $S=2$, $L=3$, and $g_j^h=1/3$ for all $h$ and $j$. The coordinates $w_j^h$ are chosen as illustrated with the pluses in the figure. In this case, we find that $\rho(z_j)$ is nonzero in two regions that are narrower in one direction and about the same size in the other direction compared to the quasihole shapes in Fig.\ \ref{fig:shape}(a). The result is hence squeezing. We find that $-\sum_{j\in B}\rho(z_j)\approx 0.5$, where the set $B$ consists of the labels of all the lattice sites inside one of the dashed ellipses. The squeezing is also seen in Fig.\ \ref{fig:shape}(d), where we show $\rho(z_j)$ along a line parallel to the real axis. By making different choices of $g_j^{h}$ and $w_j^{h}$, we can obtain different shapes.

\subsection{Braiding statistics}

We now investigate the braiding properties of the shaped objects. The computations done here generalize the computations for $L=1$ in \cite{anne2015,anne2018}. When we take the $r$th shaped object around a closed path, the wavefunction transforms as $|\psi_{q,S}\rangle \to Me^{i\theta_r}|\psi_{q,S}\rangle$, where $M$ is the monodromy, which is determined by the analytical continuation properties of the wavefunction, and
\begin{align}\label{BP}
\theta_r&=i\sum^{L}_{h=1}\oint_{c_h}\langle \psi|\frac{\partial \psi}{\partial w_{r}^{h}}\rangle dw_{r}^{h}+\textrm{c.c.}\\
&=\frac{i}{2}\sum_{h=1}^{L}\sum_{k=1}^Ng_r^{h}\oint_{c_h}\frac{\langle n_k\rangle}{w_r^h-z_k}dw_r^{h}+\textrm{c.c.}\nonumber
\end{align}
is the Berry phase. Here, we denote the path that $w_r^h$ follows by $c_h$.

The phase $\theta_r$ contains contributions both from statistical phases due to braiding and from the Aharonov-Bohm phase acquired from the background magnetic flux encircled by the quasihole. Here, we want to isolate the statistical phase when the $r$th anyon encircles the $v$th anyon. The quantity of interest is hence
\begin{multline}
\theta_{\textrm{in}}-\theta_{\textrm{out}}=\\
\frac{i}{2}\sum_{h=1}^{L}\sum_{k=1}^Ng_r^{h}\oint_{c_h}\frac{\langle n_k\rangle_{\textrm{in}}-\langle n_k\rangle_{\textrm{out}}}{w_r^h-z_k}dw_r^{h}+\textrm{c.c.},
\end{multline}
where $\theta_{\textrm{in}}$ ($\langle n_k\rangle_{\textrm{in}}$) is the Berry phase (particle density) when the $v$th anyon is well inside all the $c_h$, and $\theta_{\textrm{out}}$ ($\langle n_k\rangle_{\textrm{out}}$) is the Berry phase (particle density) when the $v$th anyon is well outside all the $c_h$. All other anyons in the system stay at fixed positions to avoid additional contributions to the Berry phase.

The density difference $\langle n_k\rangle_{\textrm{in}}-\langle n_k\rangle_{\textrm{out}}$ is nonzero only in the proximity of the two possible positions of the $v$th anyon and is hence independent of $w_r^h$. This facilitates to move the density difference outside the integral. The remaining integral $\oint_{c_h}\frac{1}{w_r^h-z_k}dw_r^{h}$ is $2\pi i$ whenever $z_k$ is inside $c_h$. Due to the assumption that the $v$th anyon is well inside or well outside $c_h$, it follows that $-\sum_{k\textrm{ inside }c_h}(\langle n_k\rangle_{\textrm{in}}-\langle n_k\rangle_{\textrm{out}})=p_v/q$ is the charge of the $v$th anyon. Utilizing further that $\sum_h g_r^h=p_r$, we conclude
\begin{equation}
\theta_{\textrm{in}}-\theta_{\textrm{out}}=2\pi p_r p_v /q.
\end{equation}
The result is hence independent of the number of weights $L$ as long as the anyons are well separated and all the weights belonging to one anyon are moved around all the weights of the other anyon.

Looking at the monodromy, a difficulty immediately arises. In the normal Laughlin state, there is a trivial contribution to the monodromy, when an anyon encircles a particle, but this is not necessarily the case here. If the weight at $w_r^h$ encircles the lattice site at $z_j$, the contribution to the monodromy is $e^{2\pi i g_r^h n_j}$, and this is nontrivial when $g_r^h$ is not an integer. In order to get Laughlin type physics, we hence need that all such factors combine to a trivial phase factor. This can be achieved by putting a restriction on how the anyons are moved. Specifically, if we require that all $c_h$ encircle the same set of lattice sites, then we get the contribution $e^{2\pi i \sum_h g_r^h n_j}=1$ for each lattice site inside the paths.

The conclusion is hence that we obtain the same braiding properties as for the normal Laughlin state as long as all the weights of an anyon encircle all the weights of another anyon and the closed paths followed by the weights belonging to an anyon all encircle the same lattice sites. This still leaves a considerable amount of freedom to shape the anyons.

\section{Squeezing and braiding anyons in the Kapit-Mueller model and in an interacting Hofstadter model}\label{sec:opt}

We next investigate anyons in the Kapit-Mueller model \cite{kapit2010} and in an interacting Hofstadter model \cite{Sorensen,Hafezi}. We choose open boundary conditions, since this is the most relevant case for experiments. We first show that one can squeeze the anyons by adding an optimized potential, and in this way it is possible to avoid significant overlap between the charge distributions of the anyons even for the small system sizes considered. We then braid the squeezed anyons and find that the Berry phase is closer to the ideal value than it is for the case without squeezing. We also demonstrate robustness with respect to small errors in the optimized potentials and estimate the time needed to reach the adiabatic limit.

\subsection{Model}

We consider hardcore bosons on a two-dimensional square lattice with $N$ sites and open boundary conditions. The hardcore bosons are allowed to hop between lattice sites, and the phases of the hopping terms correspond to a uniform magnetic field perpendicular to the plane. We study two models. The first one is the Kapit-Mueller model at half filling \cite{kapit2010} with Hamiltonian																			
\begin{equation}\label{eq:hopping}
H_0=\sum_{j\neq k} t(z_j,z_k) a^{\dagger}_j a_k + \mathrm{H.c.},
\end{equation}
where $a_j$ is the operator that annihilates a hardcore boson on the lattice site at the position $z_j$. Note that $H_0$ conserves the number of particles $\sum_i n_i$, where $n_i = a_i^\dag a_i$ is the number operator acting on site $i$. We choose the lattice spacing to be unity, and take the origin of the coordinate system to be the center of the lattice. The coefficient for the hopping from the site at $z_k$ to the site at $z_j$ is then
\begin{multline}
t(z_j,z_k)=
(-1)^{\real(z_j-z_k)+\imag(z_j-z_k)+\real(z_j-z_k)\imag(z_j-z_k)}\\
\times t_0 e^{-(\pi/2)(1-\phi)|z_j-z_k|^2}
e^{-i \pi \phi [\real(z_j)+\real(z_k)]\imag(z_j-z_k)},
\end{multline}
where $\phi$ is the number of magnetic flux units per site. We choose $t_0=1$ as the energy unit. When there are no anyons in the system, half filling means that the number of particles $\sum_i n_i$ is half the number of flux units $N\phi$. When pinning potentials are inserted to trap $S$ anyons, we should instead choose the number of particles such that $\sum_i n_i + S/2=N\phi/2$. Note that $t(z_j,z_k)$ decays as a Gaussian with distance between the two sites. If we only allow hopping between nearest neighbor sites, i.e.\ $\sum_{j\neq k}\mapsto \sum_{\langle j , k\rangle}$, then the resulting Hamiltonian is the interacting Hofstadter model \cite{Sorensen,Hafezi}. This is the second model we study.

Both models are in the same topological phase as the bosonic lattice Laughlin state with $q=2$ for appropriate choices of the parameters. For the numerical computations below, we take an $N=6 \times 6$ lattice with magnetic flux density $\phi=1/6$, and we have either three particles and no quasiholes or two particles and two quasiholes in the system. Below, we describe how we create quasiholes and braid them.

\subsection{Anyon squeezing and optimization}

\begin{figure}
\includegraphics[width=\columnwidth,trim=5mm 4mm 4mm 5mm]{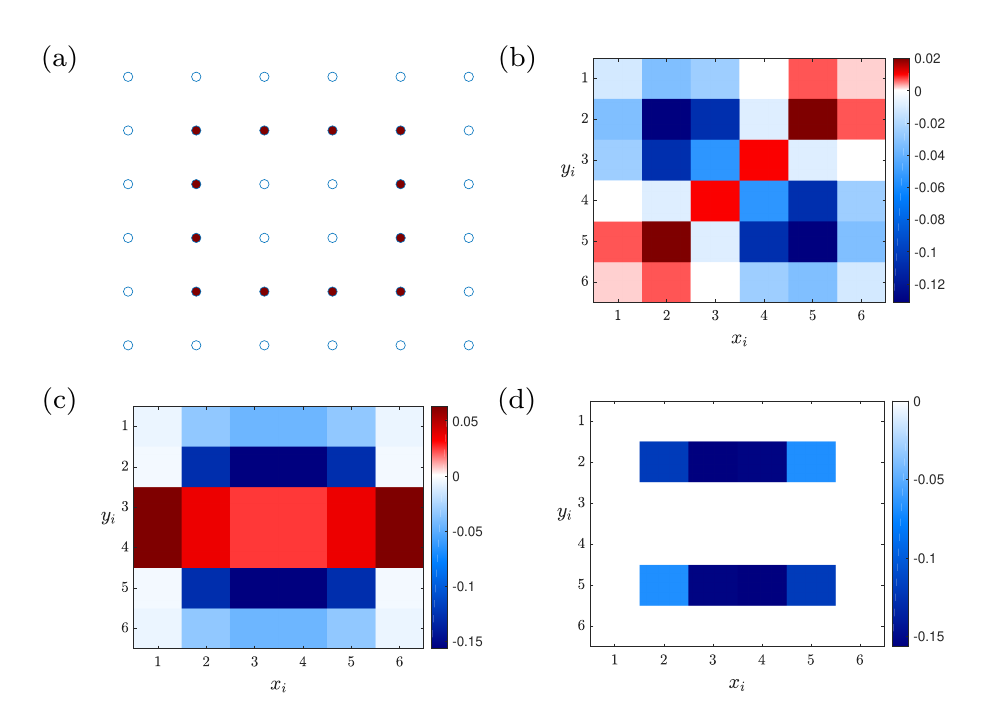}
\caption{(a) The path we choose to exchange the two quasiholes in the clockwise direction. (b) The density difference for the case of two trapping potentials at the sites $(2,2)$ and $(5,5)$. [We here label the lattice sites by $(x_i,y_i)$, where $z_i=x_i-3.5-i(y_i-3.5)$.] It is seen that the two quasiholes trapped by the potentials overlap. (c) The density difference for the case of trapping potentials on all the sites with $x_i=2,3,4,5$ and $y_i=2,5$. In this case, there is again a significant overlap between the two anyons trapped by the potentials. (d) The density difference for the case, where we optimize the potential to localize the quasiholes on the eight sites (see also the movie of braiding of the squeezed anyons). The sum of the absolute values of the charge on all the other sites is negligible with $F\sim O(10^{-15})$.}
\label{fig:compare}
\end{figure}

In Ref.\ \cite{kapit2012}, Kapit \textit{et al}.\ considered a scenario in which quasiholes are trapped by local potentials. One can pin quasiholes at the sites $a$ and $b$ by adding the term $H_p=Vn_a+Vn_b$ to the Hamiltonian described in Eq.\ \eqref{eq:hopping}, where $V$ is a positive potential strong enough to trap quasiholes. Let $\langle n_i\rangle_{H_0+H_p}$ be the expectation value of $n_i$ in the ground state of the Hamiltonian with trapping potentials $H_0+H_p$ when there are two particles and two quasiholes in the system, and let $\langle n_i\rangle_{H_0}$ be the expectation value of $n_i$ in the ground state of $H_0$ when there are three particles and no quasiholes in the system. The density difference
\begin{equation}\label{densdiffH}
\rho(z_i)= \langle n_i\rangle_{H_0+H_p} - \langle n_i\rangle_{H_0}
\end{equation}
quantifies how much the presence of the anyons alter the density, and $-\rho(z_i)$ is the charge distribution of the anyons. In the ideal case of well separated anyons, the density difference is zero everywhere, except in the vicinity of the anyons, and each anyon appears as one half particle missing on average in a local region. In Fig.\ \ref{fig:compare}(b), we show the density difference when the trapping potentials are located at the sites $(2,2)$ and $(5,5)$. Notice that the overlap between the two anyons is large compared to the absolute value of the charge at the location of the trapping potentials. One could alternatively think of having several potentials in a row as in Fig.\ \ref{fig:compare}(c), but also in this case there is a large overlap between the anyons.

Here, we show that one can avoid the overlap between the charge distributions of the two anyons by choosing the potentials appropriately, and we use optimal control to find the appropriate potentials. We apply the method to both the Kapit-Mueller and the interacting Hofstadter models. For each of the anyons we choose four consecutive sites on which we want to localize the anyon. We choose the potentials on these sites to be $(1-\lambda) V_0$, $V_0$, $V_0$, and $\lambda V_0$, respectively, where $\lambda \in [0,1]$. Later, we shall use the parameter $\lambda$ to move the anyons during the braiding operation. In addition, we introduce auxiliary potentials $V_i^{\mathrm{aux}}$ on all the other sites $i_{\mathrm{aux}}$ to achieve screening. By tuning the values of $V_i^{\mathrm{aux}}$, using optimal control theory, we are able to localize each of the anyons to very high precision on the chosen sites. We study the ground state of the Hamiltonian $H=H_0+H_p$, where $H_0$ is given by Eq.\ \eqref{eq:hopping} and the term $H_p=\sum_j V_j a_j^{\dagger} a_j$ includes the potential on all the sites.

We choose the fitness function
\begin{equation}\label{eq:fitness}
F(V_0,V_i^{\mathrm{aux}})=\sum_{i\in i_{\mathrm{aux}}} |\rho(z_i)|
\end{equation}
for the optimization to be the sum of the absolute value of the density difference over all the auxiliary sites. Note that $F$ is zero, when the anyons are perfectly localized on the chosen sites. We adopt the covariance matrix adaptation evolution strategy (CMA-ES) algorithm to minimize $F$. The CMA-ES belongs to the class of evolutionary algorithms and is a stochastic, derivative-free algorithm for global optimization. It is fast, robust and one of the most popular global optimization algorithms. See Ref.~\cite{hansen2006} for further details. We restrict the values of the auxiliary potentials to be within the interval $V_i^{\mathrm{aux}} \in [-\epsilon, \epsilon]$, and their values within this window are determined by the optimization algorithm. The hyper parameters $(V_0, \epsilon)$ are chosen empirically, but we observe that the optimization is not sensitive to the values of $V_0$ and $\epsilon$, as long as $V_0$ and $\epsilon$ are larger than certain threshold values. After the optimization the fitness function is very small with $F\sim O(10^{-15})$, and each anyon is very well localized within the desired four sites. See Fig.\ \ref{fig:compare}(d) for an illustration. In general, the optimized values of $V_i^{\mathrm{aux}}$ are smaller than $V_0$.

\begin{figure}
\includegraphics[width=0.8\columnwidth,trim=3mm 9mm 3mm 8mm]{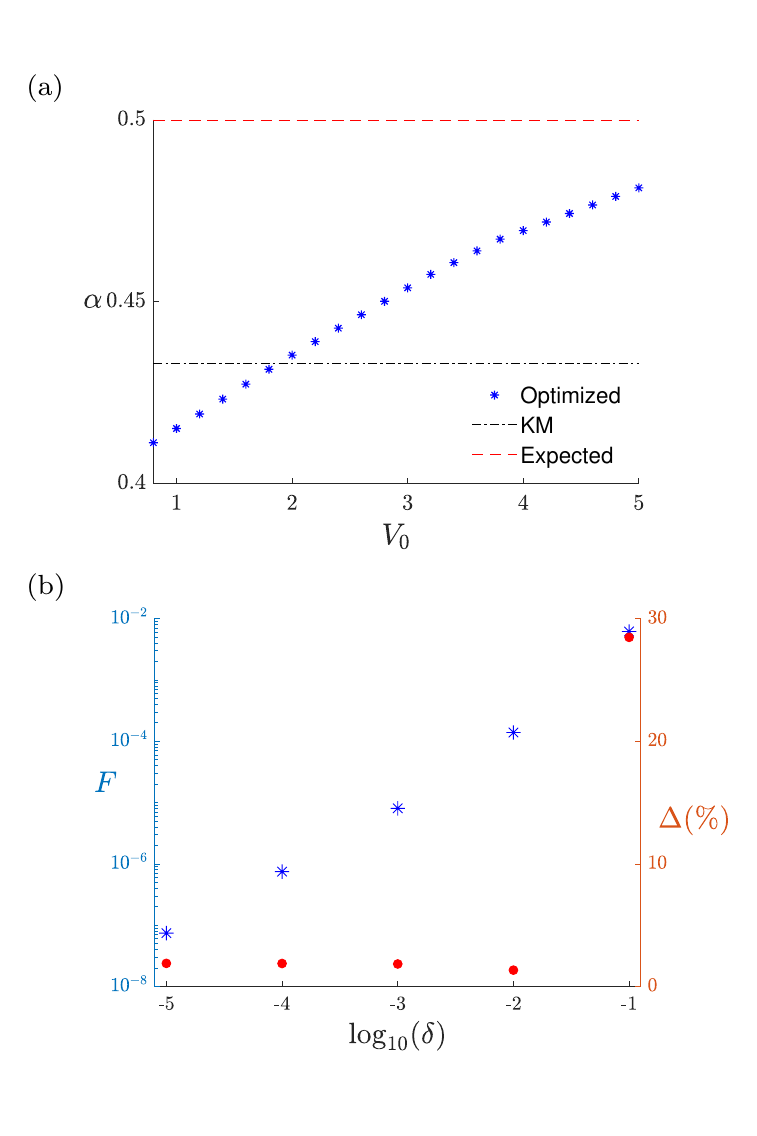}
\caption{(a) The Berry phase in units of $\pi$ versus the strength of the trapping potentials $V_0$. The blue stars are for the squeezed anyons in the Kapit-Mueller model. The red dash-dotted line denotes the expected value of the Berry phase for well separated anyons. The black dashed line marks the best value obtained in Table \ref{tab:compare} for the Kapit-Mueller model without optimization. (b) The robustness for the case with optimization against the strength of an error introduced in the potentials (see text for details). We display the values of the fitness function $F$ (blue stars) in the left axis, as well as the relative error $\Delta$ (red dots, displayed as percentages) in the right axis, once the error $\delta$ is introduced.}
\label{fig:fit_robust}
\end{figure}

\begin{table*}
\begin{tabular} {c c c c c}
\multicolumn{5}{c}{Without optimization (KM/IH)} \\
\hline
\hline
$N_\mathrm{steps}$ & $V_0=1$ & $V_0=10$ & $V_0=10^2$ & $V_0=10^3$ \\
\hline
5 & $0.2805/ 0.3173$\quad  & $0.4088/0.4476$ \quad& $0.4307/0.4685$ \quad & $\textbf{0.4329}/\textbf{0.4706}$\\
10 & 0.2815/0.3186 & 0.4098/0.4478 & 0.4308/0.4684 & $\textbf{0.4329}/\textbf{0.4706}$\\
$10^2$ & 0.2819/0.3190 & 0.4096/0.4459 & 0.4304/0.4661 & 0.4329/0.4703 \\
$10^3$ & 0.2819/0.3191 & 0.4095/0.4458 & 0.4296/0.4636 & 0.4324/0.4678 \\
\hline
\hline
\end{tabular}

\vspace{3mm}

\begin{tabular}{c c c}
\multicolumn{3}{c}{With optimization} \\
\hline
\hline
$N_\mathrm{steps}$ & $V_0=5$ & $V_0=7$\\
\hline
$4$& \textbf{0.4873}\;(KM) & \textbf{0.5062}\;(IH)\\
\hline
\hline
\end{tabular}

\caption{Berry phases $\alpha$ in units of $\pi$ obtained without (top panel) and with (bottom panel) optimization for the exchange of two quasiholes in the Kapit-Mueller (KM) model and in the interacting Hofstadter (IH) model with open boundary conditions. The values of $\alpha$ obtained from the two methods that are closest to the ideal value $1/2$ are indicated with bold text.}\label{tab:compare}
\end{table*}

\subsection{Adiabatic exchange}

To exchange the anyons, we need the potentials to vary in time. For the case without optimization considered in \cite{kapit2012}, the potentials were moved between two sites by linearly increasing the strength at one site and linearly decreasing the strength at the other site. In our case, we move the chain of trapping potentials forward by one site by linearly increasing $\lambda$ from $0$ to $1$. We move both the anyons simultaneously in the clockwise direction. We discretize this process into $N_{\mathrm{steps}}$ steps. As described before, we also include auxiliary potentials on all the other sites, and we optimize these for each value of $\lambda$. By repeating this procedure, we make a complete exchange of the two anyons.

We compute the Berry phase factor $e^{i\pi\alpha}$ for the exchange both with and without optimization. For the case without optimization, we start out with the trapping potentials at the lattice sites (2,2) and (5,5). For the case with optimization, we start from the situation depicted in Fig.\ \ref{fig:compare}(d). Here, we assume the adiabatic limit and follow the method in Ref.\ \cite{kapit2012} to compute the Berry phase factor. Specifically, we compute the ground state at each step by exact diagonalization, and we fix the phase of the ground state by requiring that the overlap between the wavefunction at the current step and the wavefunction at the previous step is real. The Berry phase can then be read off by comparing the phase of the final state to the phase of the initial state. The expected value for the case of well separated anyons is $\alpha=1/2$ and hence $e^{i\pi\alpha}=i$.

In Table~\ref{tab:compare}, we compare the cases with and without optimization for both the Kapit-Mueller model and the interacting Hofstadter model. We find that a small value of $N_{\mathrm{steps}}$ is enough to reach the regime, in which the Berry phase $\alpha$ is not sensitive to the precise choice of $N_\mathrm{steps}$. For the case without optimization, we find that a high strength of the potential is required ($V_0\geq 100$) to obtain reasonably good results. For the case with optimization, however, a quite weak potential $V_0<10$ is sufficient. In addition, the Berry phases obtained with optimization are much closer to the ideal value than the ones obtained without optimization. This can also be seen in Fig.\ \ref{fig:fit_robust}(a), which shows the effect of the strength $V_0$ on the numerical value of $e^{i\pi\alpha}$ for the case with optimization. In general, the larger $V_0$ is, the better the $e^{i\pi\alpha}$ is. However, the numerical result saturates when $V_0$ exceeds a threshold value and thus strong $V_0$ is not necessary.

\subsection{Robustness}

In an experiment, the desired potentials on different lattice sites may not be exactly achieved, and we therefore test how robust the scheme is against possible errors in the auxiliary potentials as shown in Fig.\ \ref{fig:fit_robust}(b). We consider potentials $V_i^{'}=V_i^{\mathrm{opt}}(1+\delta)$ that are slightly perturbed away from the optimized potentials $V_i^{\mathrm{opt}}$ due to an error $\delta$ coming from uncertainties in the operations in the experiment. First, we show the robustness of our method with respect to screening of the anyons by plotting the fitness $F$ as a function of the size of the error $\delta$ introduced. We observe that when the error is reasonably small, say $\delta<10^{-2}$, the fitness function is quite small $F<10^{-4}$, thus the anyons are still very well screened. Secondly, we demonstrate the robustness with respect to the relative error of the Berry phase once $\delta$ is introduced. The relative error of the Berry phase is defined as
\begin{equation}
\Delta=\frac{\left|\mathrm{Im}(e^{i \pi\alpha_{\delta}})- \mathrm{Im}(e^{i \pi\alpha})\right|}{\mathrm{Im}(e^{i \pi\alpha})},
\end{equation}
where $\alpha_{\delta}$ is the Berry phase obtained with the error. Notice that the relative error is small $\Delta\sim O(10^{-2})$ when the error is reasonably small $\delta \leq 10^{-2}$.

\subsection{Braiding within a finite time}

\begin{figure}
\includegraphics[width=\columnwidth]{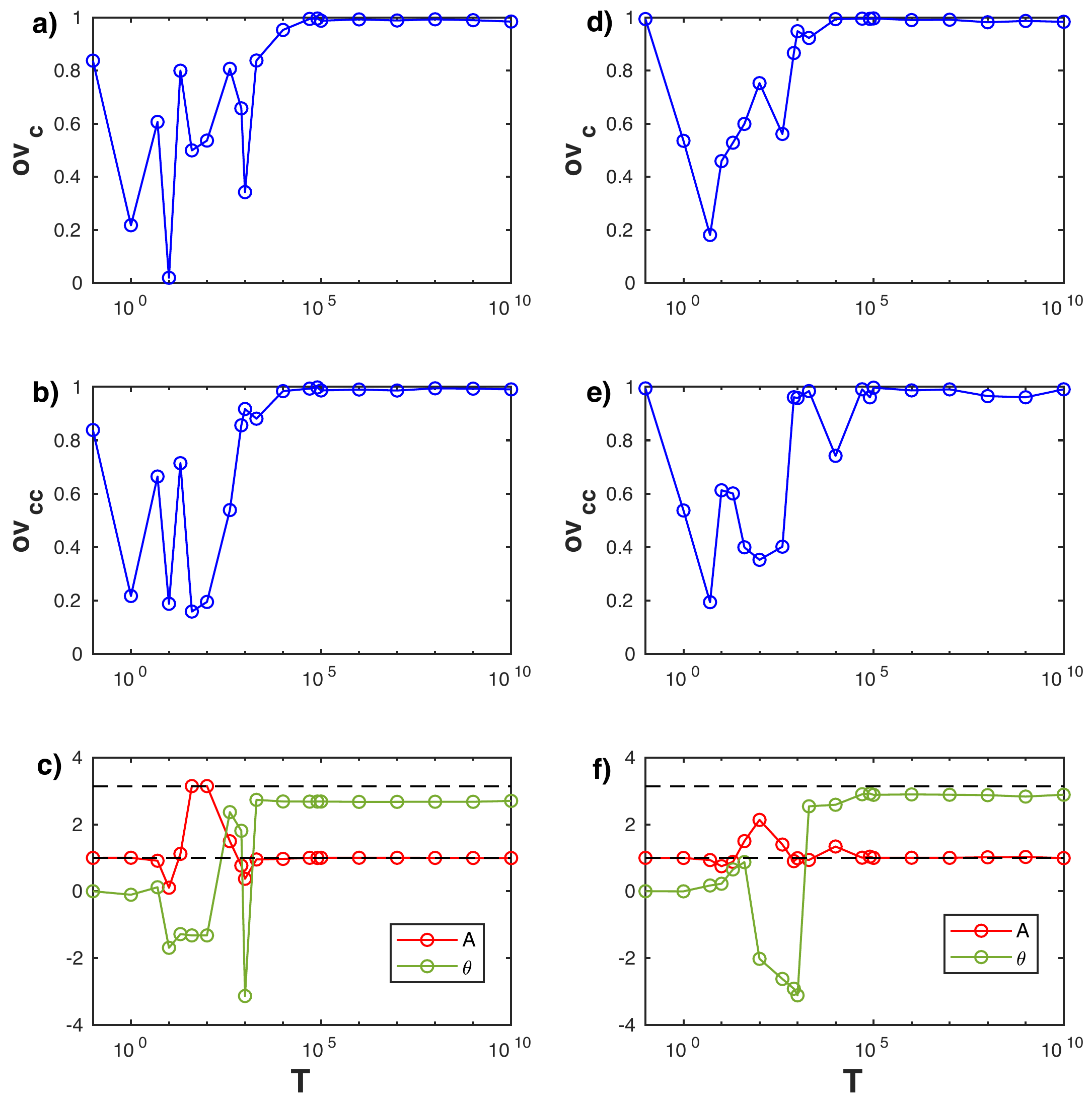}
\caption{Braiding results for the exchange of two anyons in the Kapit-Mueller model. The left (right) hand side of the figure is without (with) optimization. (a,d) The overlap $\textrm{ov}_\textrm{c}=|\langle\psi(0)|\psi(T)\rangle_{\textrm{clockwise}}|$ between the initial and final states when the anyons are exchanged in the clockwise direction as a function of the duration $T$ of the exchange. (b,e) The same quantity, but for a counterclockwise exchange. (c,f) Amplitude $A$ and phase $\theta$ of the ratio $R$ of the overall phases computed for the clockwise and counterclockwise cases. The dashed lines show the ideal values $A=1$ and $\theta=\pi$.}\label{fig:time}
\end{figure}

The computation above to determine the Berry phase assumes that the anyons are exchanged infinitely slowly to ensure that the ground state returns to itself with no mixing with the excited states. It is, however, also relevant to know how slowly the anyons should be moved to be close to adiabaticity, and we therefore now consider the outcome when the braiding is completed within a fixed time $T$.

Ideally one would start with an initial wavefunction $|\psi(0)\rangle$ and achieve the wavefunction at time $t$ through the expression
\begin{equation}
|\psi(t)\rangle=U(t)|\psi(0)\rangle,
\end{equation}
where $U(t)=\mathcal{T}\exp[-i\int_{0}^{t}H(t')dt']$ is the time evolution operator, $\mathcal{T}$ stands for time ordering, and $H(t)=H_0+H_p(t)$ is the time-dependent Hamiltonian. Numerically, however, one must split the unitary operator into a finite number of steps $\mathcal{N}$, and hence we use the relation
\begin{equation}
U(t)=\lim_{\mathcal{N}\rightarrow \infty} \prod_{j=0}^{\mathcal{N}-1}e^{-iH(j\Delta t)\Delta t}.
\end{equation}
In the above expression, $\Delta t=t/\mathcal{N}$ and we consider large $\mathcal{N}$ to ensure convergence.

For the case without optimization, we move the potential of strength $V$ from say site $a$ to site $b$ in $\mathcal{N}$ steps using
\begin{equation}
H_p(j\Delta t)=\left[\cos^2\left(\frac{j\pi}{2\mathcal{N}}\right)\,n_a
+\sin^2\left(\frac{j\pi}{2\mathcal{N}}\right)\,n_{b}\right]V
\end{equation}
where $j\in\{0,1,\ldots,\mathcal{N}-1\}$. We compute this for the Kapit-Mueller model and use the same braiding path as shown in Fig.\ \ref{fig:compare}(a).

For the case with optimization, we reuse the optimized potentials computed for the Kapit-Mueller model with $N_{\textrm{steps}}=4$. Here, we need additional steps to ensure convergence, and we hence use $\mathcal{N}N_{\textrm{steps}}$ steps to move the anyons one lattice spacing. Instead of doing an optimization for each of these smaller steps, we smoothly interpolate between the potentials already computed for two successive steps $i$ and $i+1$ through
\begin{multline}
H_p(i\mathcal{N}\Delta t+j\Delta t)=\\
\cos^2\left(\frac{j\pi}{2\mathcal{N}}\right)\,H^{i}_p
+\sin^2\left(\frac{j\pi}{2\mathcal{N}}\right)\,H^{i+1}_{p},
\end{multline}
where $i\in\{0,1,\ldots,N_{\textrm{steps}}-1\}$, $j\in\{0,1,\ldots,\mathcal{N}-1\}$ and $H^{i}_p$ are the optimized potentials computed for the $i$th step. By this procedure, we achieve a total of $4\mathcal{N}$ steps quite easily which would otherwise require a huge numerical effort if we had to run the optimization algorithm for each of the $4\mathcal{N}$ steps.

We repeat this procedure until the anyons have been exchanged, and we let $T$ denote the total time used for the exchange. When $T$ is sufficiently large, the process is adiabatic, and the phase acquired during the exchange is $\langle\psi(0)|\psi(T)\rangle$. This phase, however, has contributions from both the dynamical phase and the statistical phase due to the exchange of the anyons. The dynamical phase is prone to numerical error accumulated due to variations in the instantaneous energies at each point along the braiding path. We also note that exchanging the two anyons clockwise and counterclockwise results in the same dynamical phase and hence also the same numerical error. We can hence get rid of the dynamical phase by computing the ratio
\begin{equation}
R=\frac{\langle\psi(0)|\psi(T)\rangle_{\textrm{clockwise}}}
{\langle\psi(0)|\psi(T)\rangle_{\textrm{counterclockwise}}}.
\end{equation}
In general, $R=A\exp{(i\theta)}$ and for the ideal case with well-separated anyons, $A=1$ and $\theta=\pi$.

We compute $\theta$, $A$, and the overlap between the initial and final wavefunctions for the cases with and without optimization for different total times $T$ for the Kapit-Mueller model as shown in Fig.\ \ref{fig:time}. The time required to achieve adiabaticity for the case with optimization ($T_{\textrm{opt}}$) is slightly larger compared to the case without optimization ($T_{\textrm{normal}}$), and roughly $T_{\textrm{opt}}\approx 5T_{\textrm{normal}}$. This is possibly due to the smaller energy gap to the first excited state for the case with optimization compared to the case without optimization. We see that the braiding phase is better for the case with optimization due to the better screening of the anyons and we believe a strict optimization at each step will result in a more accurate braiding phase.

\section{Conclusion} \label{sec:conclusion}

We have shown that one can use optimal control techniques to squeeze anyons and in this way braid them within a smaller area. This is an advantage, since smaller systems are often easier to deal with, and it allows a larger number of operations to be done per area. We find that the squeezed anyons need to move at a slower speed to be close to the adiabatic limit (by a factor of about five for the considered example). We note, however, that to get better results for the Berry phase for the normal anyons, one would need to separate the anyons further, which would also increase the time needed for the braiding.

An important property of anyon braiding is the topological robustness against small, local disturbances. In the scheme used here, we only optimize the shape of the anyons and not the Berry phase itself, and we rely on adiabatic time evolution to obtain the Berry phase. In this way the topological robustness is maintained.

Using a relatively simple physical quantity for the optimization can also be an advantage for experiments. Even if one does not know the model for the considered system exactly, one can experimentally find the correct potential to squeeze the anyons, if one can measure the density.

We here considered Abelian anyons. It would also be interested to use similar ideas to squeeze and braid non-Abelian anyons.

\begin{acknowledgments}
The authors thank Jacob F. Sherson and Jens Jakob S{\o}rensen for discussions on optimal control theory. This work was in part supported by the Independent Research Fund Denmark under grant number 8049-00074B.
\end{acknowledgments}


%

\end{document}